\documentclass[onecolumn]{aa}
\usepackage{graphicx}
\usepackage{txfonts}

\newcommand{\beq}{\begin{equation}}
\newcommand{\beqa}{\begin{eqnarray}}
\newcommand{\eeq}{\end{equation}}
\newcommand{\eeqa}{\end{eqnarray}}
\newcommand{\etal}{{\it et al. }}

\newcommand{\lsim}{\la}
\newcommand{\gsim}{\ga}

\newcommand{\bfxi}{\mbox{\boldmath{$\xi$}}}
\newcommand{\bfeta}{\mbox{\boldmath{$\eta$}}}
\begin{document}
   \title{Quasi-geometrical Optics Approximation in
 Gravitational Lensing}

   \subtitle{}

   \author{R. Takahashi}

   \offprints{R. Takahashi}


   \institute{Division of Theoretical Astrophysics, 
              National Astronomical Observatory, Mitaka,
              Tokyo 181-8588, Japan \\
              \email{takahasi@th.nao.ac.jp}
             }

   \date{Received; accepted}

   \abstract{The gravitational lensing of gravitational waves should
 be treated in the wave optics instead of the geometrical optics
 when the wave length $\lambda$ of the gravitational waves
 is larger than the Schwarzschild radius of the lens mass $M$.
 The wave optics is based on the diffraction integral which represents
 the amplification of the wave amplitude by lensing.
 We study the asymptotic expansion of the diffraction integral in
 the powers of the wave length $\lambda$.
 The first term, arising from the short wavelength limit $\lambda \to
 0$, corresponds to the geometrical optics limit.
 The second term, being of the order of $\lambda/M$, is the leading
 correction term arising from the diffraction effect.
 By analysing this correction term, we find that (1) the lensing
 magnification $\mu$ is modified to $\mu ~(1+\delta)$, where $\delta$
 is of the order of $(\lambda/M)^2$, and (2) if the lens has cuspy
 (or singular) density profile at the center $\rho(r) \propto
 r^{-\alpha}$ ($0 < \alpha \leq 2$), the diffracted image is formed
 at the lens center with the magnification $\mu \sim
 (\lambda/M)^{3-\alpha}$.
   }

   \maketitle
%

\section{Introduction}

The gravitational lensing of light is usually treated in the geometrical
 optics approximation, which is valid in almost all observational 
 situations since the wave length of light is much
 smaller than typical scales of astrophysical lens objects.
However for the gravitational lensing of gravitational waves,
 the wavelength is long so that the geometrical optics approximation is 
 not valid in some cases.  
As shown by several authors (Ohanian 1974, Bliokh \& Minakov 1975,
 Bontz \& Haugan 1981, Thorne 1983, Deguchi \& Watson 1986),
 if the wavelength $\lambda$ is larger than the Schwarzschild radius
 of the lens mass $M$, the diffraction effect becomes important.
Thus, the diffraction effect is important for the lens mass smaller
 than $10^8 M_{\odot} ~( {\lambda}/{1 {\mbox{AU}}} )$,
 where $1$ AU is the wavelength 
 for the planed laser interferometer space antenna
  (LISA: Bender \etal \cite{b00})

From the above discussion, for $\lambda \gsim M$ the diffraction effect
 is important and the magnification is small (the wavelength is so long
 that the wave does not feel the existence of the lens), and
 for $\lambda \ll M$ the geometrical optics approximation is valid.  
In this paper, we consider the case for $\lambda \lsim M$, i.e.  
 the quasi-geometrical optics approximation which is 
 the geometrical optics including corrections
 arising from the effects of the finite wavelength.
We can obtain these correction terms by an asymptotic expansion of
 the diffraction integral in powers of the wavelength
 $\lambda$.\footnote{The asymptotic
 expansion of the diffraction integral has been studied in optics.
 See the following text books for detailed discussion:
 Kline \& Kay (1965), Ch.XII; Mandel \& Wolf (1995), Ch.3.3;
 Born \& Wolf (1997), App.III, and references therein.}\
The diffraction integral represents the amplification of the
 wave amplitude by lensing in the wave optics.
It is important to derive the correction terms
 for the following two reasons:
(1) calculations in the wave optics are based on the diffraction
 integral, but it is time consuming to numerically calculate this
 integral especially for high frequency
 (see e.g. Ulmer \& Goodman \cite{ug95}).
 Hence, it is a great saving of computing time to use the analytical
 expressions.
(2) We can understand clearly the difference between the
 wave optics and the geometrical optics (i.e. the diffraction effect).

This paper is organized as follows:
In \S 2 we briefly discuss the wave optics in gravitational lensing of
 gravitational waves. 
In \S 3 we show that in the short wavelength limit $\lambda \to 0$,
 the wave optics is reduced to the geometrical optics limit. 
In \S 4 we expand the diffraction integral in powers of the 
 wavelength $\lambda$, and derive the leading correction terms arising
 from the effect of the finite wavelength. 
In \S 5 we apply the quasi-geometrical optics approximation to the case
 of the simple lens models (the point mass lens, SIS lens, isothermal
 sphere with a finite core lens, and the NFW lens).
Section 6 is devoted to summary and discussions.
We use units of $c=G=1$.


\section{Wave Optics in Gravitational Lensing}


We consider gravitational waves propagating near a lens object
 under the thin lens approximation in which the gravitational wave
 is scattered only on the thin lens plane.  
The gravitational wave amplitude is magnified by the amplification
 factor $F$ which is given by the diffraction integral as
 (Schneider \etal \cite{sef92}),
\beq
F(\omega,\bfeta)=\frac{D_S}{D_L D_{LS}} \frac{\omega}{2 \pi i}
 \int d^2 \xi ~\exp\left[ i \omega t_d(\bfxi,\bfeta) \right],
\label{ampf1} 
\eeq  
where $\omega$ is the frequency of the wave, 
 $\bfxi$ is the impact parameter in the lens plane,
 $\bfeta$ is the source position in the source plane.
$D_L, D_S$ and $D_{LS}$ are the distances to the lens, the source
 and from the source to the lens, respectively.
The time delay $t_d$ from the source position
 $\bfeta$ through $\bfxi$ is given by,
\beq
 t_d(\bfxi,\bfeta)=\frac{D_L D_S}{2 D_{LS}}
 \left( \frac{\bfxi}{D_L} - \frac{\bfeta}{D_S} \right)^2 -
 \hat{\psi}(\bfxi) + \hat{\phi}_m(\bfeta ).
\label{td1}
\eeq 
The deflection potential $\hat{\psi}(\bfxi)$ is determined by, 
 $\nabla_{\xi}^2 \hat{\psi} = 8 \pi \Sigma$,
where $\Sigma(\bfxi)$ is the surface mass density of the lens.
$\hat{\phi}_m$ is the additional phase in $F$, and we choose
 $\hat{\phi}_m$ so that the minimum value of the time delay is zero.
 $F$ is normalized such that
 $|F|=1$ in no lens limit ($\hat{\psi}=0$). 

It is useful to rewrite the amplification factor $F$ in terms of
 dimensionless quantities.
We introduce $\xi_0$ as the normalization constant
 of the length in the lens plane.
The impact parameter $\bfxi$, the source position $\bfeta$, 
 the frequency $\omega$, and the time delay $t_d$ are rewritten
 in dimensionless form, 
\beq
  \mathbf{x}=\frac{\bfxi}{\xi_0}, ~~\mathbf{y}=\frac{D_L}{\xi_0 D_S}
 \bfeta, ~~w = \frac{D_S}{D_{LS} D_L} \xi_0^2 ~\omega,
 ~~ T=\frac{D_L D_{LS}}{D_S} \xi_0^{-2} t_d.
\label{xydef}
\eeq
We use the Einstein radius ($\sim (M D)^{1/2}$)
 as the arbitrary scale length $\xi_0$ for convenience.
Then, the dimensionless frequency is $w \sim M/\lambda$
 from Eq.(\ref{xydef}).
The dimensionless time delay is rewritten from Eq.(\ref{td1})
 and (\ref{xydef}) as,
\beqa
  T(\mathbf{x},\mathbf{y})  = \frac{1}{2} \left| \mathbf{x}-\mathbf{y}
  \right|^2  - \psi(\mathbf{x}) + \phi_m (\mathbf{y}),
\label{timed}
\eeqa 
where $\psi(\mathbf{x})$ and $\phi_m(\mathbf{y})$ correspond to 
 $\hat{\psi}(\bfxi)$ and $\hat{\phi}_m (\bfeta)$ in Eq.(\ref{td1}):
 $(\psi, \phi_m)=D_L D_{LS}/(D_S \xi_0^2) ~
 (\hat{\psi}, \hat{\phi}_m)$.
Using the above dimensionless quantities, the amplification factor
 is rewritten as, 
\beq
  F(w,\mathbf{y})=\frac{w}{2 \pi i} \int d^2 x \exp \left[ i w 
 T(\mathbf{x},\mathbf{y}) \right].
\label{ampf}
\eeq     
For the axially symmetric lens models, the deflection potential
 $\psi(\mathbf{x})$ depends only on $x=|\mathbf{x}|$ and 
 $F$ is rewritten as,
\beq
  F(w,y) = -i w e^{i w y^2/2} \int_0^{\infty} dx ~x ~J_0 (w x y)  
 \exp \left[ i w \left( \frac{1}{2} x^2 - \psi(x) + \phi_m(y) \right)
 \right],
\label{axiampf}
\eeq 
where $J_0$ is the Bessel function of zeroth order.
It takes a long time to calculate $F$ numerically in
 Eqs.(\ref{ampf}) and (\ref{axiampf}), because the integrand is rapidly 
 oscillating function especially for large $w$.
In the Appendix A, we present a method for numerical computation
 to shorten the computing time.

\section{Geometrical Optics Approximation}

In the geometrical optics limit ($w \gg 1$), the stationary 
 points of the $T(\mathbf{x},\mathbf{y})$ 
 contribute to the integral of Eq.(\ref{ampf}) so that the image
 positions $\mathbf{x}_j$ are determined by the lens equation,
 $\nabla_x T(\mathbf{x},\mathbf{y})=0$ or
 $\mathbf{y}=\mathbf{x}-\nabla_x \psi(\mathbf{x})$.
This is just the Fermat's principle. 
We expand $T(\mathbf{x},\mathbf{y})$ around the j-th image position
 $\mathbf{x}_j$ as,
\beq
  T(\mathbf{x},\mathbf{y}) = T_j
 + \frac{1}{2} \sum_{a,b} \partial_a \partial_b
 T(\mathbf{x}_j,\mathbf{y}) \tilde{x}_a \tilde{x}_b 
 + \mathcal{O} (\tilde{x}^3)
\label{expt}
\eeq
where $\tilde{\mathbf{x}}=\mathbf{x}-\mathbf{x}_j$,
 $T_{j}=T (\mathbf{x}_j,\mathbf{y})$, and the indices 
 a,b,... run from 1 to 2.
Inserting Eq.(\ref{expt}) to (\ref{ampf}),
 we obtain the amplification factor in the geometrical optics limit as
 (Schneider \etal \cite{sef92}; Nakamura \& Deguchi \cite{nd99}),
\beq
 F_{geo}(w,\mathbf{y})=\sum_j \left| \mu_j \right|^{1/2} 
 ~e^{i w T_{j}- i \pi n_j},
\label{fgeo}
\eeq
where the magnification of the j-th image is $\mu_j=1/\det \left( 
 \partial \mathbf{y} / \partial \mathbf{x}_j \right)$
 and $n_j=0,1/2,1$ when $\mathbf{x}_j$ is a minimum, saddle,
 maximum point of $T (\mathbf{x},\mathbf{y})$.

\section{Quasi-geometrical Optics Approximation}

\subsection{Effect on the Magnification of the Image}

We expand the amplification factor $F(w,\mathbf{y})$ in powers of
 $1/w~(\ll 1)$ and discuss the behavior of the order of $1/w$ term.
Here, we only consider axially symmetric lens models because 
 the basic formulae are relatively simple,
 while the case of the non-axially symmetric lens models is discussed
 in the Appendix B.
We expand $T(\mathbf{x},\mathbf{y})$ in Eq.(\ref{expt}) up to
 the fourth order of $\tilde{x}$ as,
\beq
  T(\mathbf{x},\mathbf{y}) = T_j + \alpha_j \tilde{x}_1^2 
 + \beta_j \tilde{x}_2^2 
 + \frac{1}{6} \sum_{a,b,c} \partial_a \partial_b \partial_c
 T(\mathbf{x}_j,\mathbf{y}) \tilde{x}_a \tilde{x}_b \tilde{x}_c 
 + \frac{1}{24} \sum_{a,b,c,d} \partial_a \partial_b \partial_c
 \partial_d T(\mathbf{x}_j,\mathbf{y}) \tilde{x}_a \tilde{x}_b
 \tilde{x}_c \tilde{x}_d
 + \mathcal{O} (\tilde{x}^5),
\label{expt2}
\eeq
where $\alpha_j$ and $\beta_j$ are defined by,
\beq
  \alpha_j = \frac{1}{2} \left( 1- \psi_j^{\prime \prime} \right) , 
  ~~\beta_j = \frac{1}{2} \left( 1- \frac{\psi_j^{\prime}}{|x_j|} \right) 
\label{expt1-4}
\eeq
where $\psi_j^{(n)}=d^n \psi(|\mathbf{x}_j|) / d x^n$.
Inserting Eq.(\ref{expt2}) into (\ref{ampf}) and expanding Eq.(\ref{ampf})
 in powers of $1/w$, we obtain, 
\beqa
  F(w,y) &=&
 \frac{1}{2 \pi i} \sum_j e^{i w T_j} \int d^2 \mathbf{x^{\prime}}
 e^{ i \left(  \alpha_j x^{\prime 2}_1 + \beta_j x^{\prime 2}_2
 \right)} \left[ 1 + \frac{i}{6 \sqrt{w}} \sum_{a,b,c}
 \partial_a \partial_b \partial_c T(\mathbf{x}_j,\mathbf{y})
 x^{\prime}_a x^{\prime}_b x^{\prime}_c  
 + \frac{1}{w} \left\{ - \frac{1}{72} \left( 
 \sum_{a,b,c} \partial_a \partial_b \partial_c
 T(\mathbf{x}_j,\mathbf{y}) x^{\prime}_a x^{\prime}_b
 x^{\prime}_c \right)^2 \right. \right. \nonumber \\
 &&  ~+ \left. \left. \frac{i}{24} \sum_{a,b,c,d}
 \partial_a \partial_b \partial_c \partial_d T(\mathbf{x}_j,\mathbf{y})
 x^{\prime}_a x^{\prime}_b x^{\prime}_c x^{\prime}_d \right\} 
 + \mathcal{O} (w^{-3/2}) \right],
\label{ampf5}
\eeqa
where $\mathbf{x}^{\prime}=\sqrt{w} \tilde{\mathbf{x}}$.
The first term of the above equation (\ref{ampf5}) is the amplification
 factor in the geometrical optics limit $F_{geo}$ in Eq.(\ref{fgeo}).
The integral in the second term vanishes because  
 the integrand is an odd function of $x^{\prime}_a$.
The third term is the leading correction term, being proportional to
 $1/w$, arising from the diffraction effect.
Thus
 the deviation from the
 geometrical optics is of the order of $1/w \sim \lambda/M$.
Inserting $T(\mathbf{x},\mathbf{y})$ in Eq.(\ref{timed}) into
 (\ref{ampf5}), we obtain $F$ as,
\beq
   F(w,y)=\sum_j \left| \mu_j \right|^{1/2}
 \left( 1+ \frac{i}{w} \Delta_j \right) e^{i w T_{j}- i \pi n_j}
 + \mathcal{O} (w^{-2}),
\label{fdgeo}
\eeq
where 
\beq
  \Delta_j = \frac{1}{16} \left[ \frac{1}{2 \alpha_j^2} \psi_j^{(4)}
 + \frac{5}{12 \alpha_j^3} {\psi_j^{(3)}}^2  + \frac{1}{\alpha_j^2}
 \frac{\psi_j^{(3)}}{|x_j|} + \frac{\alpha_j - \beta_j}{\alpha_j
 \beta_j} \frac{1}{|x_j|^2}  \right],
\eeq
and $\Delta_j$ is a real number.
We denote $dF_m$ as the second term of Eq.(\ref{fdgeo}),
\beq
 dF_m(w,y) \equiv \frac{i}{w} \sum_j  \Delta_j 
 \left| \mu_j \right|^{1/2}  e^{i w T_{j}- i \pi n_j}.
\label{dfm}
\eeq
Since $dF_m$ is the correction term arising near the image positions,
 this term represents the corrections to the properties of these images
 such as its magnifications, and the time delays.  
We rewrite $F$ in above equation (\ref{fdgeo}) as,
 $F(w,y)=\sum_j |~\mu_j [ 1+( {\Delta}_j/w )^2 ] |^{1/2}
 e^{i w T_{j} +i \delta \varphi_j
 - i \pi n_j} + \mathcal{O} (w^{-2})$,
where $\delta \varphi_j = \arctan ({\Delta}_j/w)$.
Thus in the quasi-geometrical optics approximation, the magnification 
 $\mu_j$ is modified to $\mu_j [1+({\Delta}_j/w)^2]$, where
 $(\Delta_j/w)^2$ is of the order of $(\lambda/M)^2$.
That is, the magnification is slightly larger than that in the
 geometrical optics limit.
The phase is also changed by $\delta \varphi_j$, which is of the order of
 $\lambda/M$.


\subsection{Contributions from the Non-stationary Points}

In the previous section, we showed that the contributions to the
 diffraction integral $F$ arise
 from the stationary points (or image positions).
In this section, we discuss the contributions from the non-stationary
 points.
We denote $\mathbf{x}_{ns}$ as the non-stationary point,
 at which the condition $|\nabla_x T| \neq 0$ satisfies.
If $T(\mathbf{x},\mathbf{y})$ can be expanded at $\mathbf{x}_{ns}$,
 we obtain the series of $T$ similar to Eq.(\ref{expt}) as,
\beq
  T(\mathbf{x},\mathbf{y}) = T_{ns} + T^{\prime}_{1} \tilde{x}_1
 + T^{\prime}_{2} \tilde{x}_2 + \alpha_{ns} \tilde{x}_1^2 
 + \beta_{ns} \tilde{x}_2^2 + \mathcal{O} (\tilde{x}^3),
\label{exptns}
\eeq
where $\tilde{\mathbf{x}}=\mathbf{x}-\mathbf{x}_{ns}$,
 $T_{ns}=T(\mathbf{x}_{ns},\mathbf{y})$ and
 $T^{\prime}_{1,2} = \partial_{1,2} T(\mathbf{x}_{ns},\mathbf{y})$.
Note that either $T^{\prime}_{1}$ or $T^{\prime}_{2}$ does not vanish
 because of $|\nabla T| \neq 0$ at $\mathbf{x}_{ns}$.
$\alpha_{ns}$, and $\beta_{ns}$ are defined same as Eq.(\ref{expt1-4}),
 but at the non-stationary point $\mathbf{x}_{ns}$. 

Inserting Eq.(\ref{exptns}) to (\ref{ampf}) and expanding the integrand
 in powers of $1/w$, we obtain, 
\beq
  F(w,y) = 2 \pi \frac{e^{i w T_{ns}}}{i w} \left[
 \delta(T^{\prime}_1) \delta(T^{\prime}_2) - \frac{i}{w}
 \left( \alpha_{ns} \frac{\partial^2}{\partial T^{\prime 2}_1} +
 \beta_{ns} \frac{\partial^2}{\partial T^{\prime 2}_2} \right)
 \delta(T^{\prime}_1) \delta(T^{\prime}_2) + \mathcal{O} (w^{-2})
 \right].
\label{nsf2}
\eeq  
Thus, if both $T^{\prime}_1 \neq 0$ and $T^{\prime}_2 \neq 0$, the
 above equation vanishes.
Even if either $T^{\prime}_1=0$ or $T^{\prime}_2=0$, it is easy to
 show that $F$ vanishes.
Thus, the contributions to the amplification factor $F$ at the
 non-stationary points are negligible.  

In the above discussion, we assume that $T(\mathbf{x},\mathbf{y})$
 has the derivatives at the non-stationary point $\mathbf{x}_{ns}$
 in Eq.(\ref{exptns}).
But, if the derivatives of $T$ are not defined at $\mathbf{x}_{ns}$,
 the result in Eq.(\ref{nsf2}) should be reconsidered. 
If the lens has the cuspy (or singular) density profile at the center, 
 the derivatives of $T$ are not defined at the lens center.
We will discuss this case in the next section.

\subsection{Central Cusp of the Lens}

We consider the correction terms in the amplification
 factor $F$ arising at the central cusp of the lens.
For the inner density profile of the lens $\rho \propto r^{-\alpha}$
 ($0 < \alpha \leq 2$),\footnote{We do not consider the steeper
 profile $\alpha > 2$, since the mass at
 the lens center is infinite.}\ 
 the surface density and the deflection potential at small radius
 are given by,
\beqa
  \Sigma(\bfxi) &\propto& \xi^{-\alpha+1}   ~~~\mbox{for}~\alpha \neq 1,  
  \nonumber \\
  &\propto& \ln \xi   \hspace{0.52cm} \mbox{for}~\alpha = 1,
\eeqa  
\beqa
  \psi(\mathbf{x}) &\propto& x^{-\alpha+3} ~~~~~~\mbox{for}~\alpha \neq 1,  
  \nonumber \\
  &\propto& x^2 \ln x   \hspace{0.42cm} \mbox{for}~\alpha = 1.
\label{apsi}
\eeqa    
We note that the Taylor series of $\psi(\mathbf{x})$ around
 $\mathbf{x}=0$ like Eq.(\ref{exptns}) cannot be obtained 
 from the above equation (\ref{apsi}).
For example, in the case of $\alpha=2$, the deflection potential is
 $\psi \propto x$, but the derivative of 
 $|\mathbf{x}|$ is discontinuous at $\mathbf{x}=0$.
Hence we use $\psi$ in Eq.(\ref{apsi}) directly, not the
 Taylor series.
Let us calculate the correction terms in the amplification factor
 arised from the lens center for following cases;
 $\alpha=2,1$, and the others.

\paragraph{Case of $\mbox{\boldmath{$\alpha =2$}}$}  

If the inner density profile is $\rho \propto r^{-2}$ (e.g. the singular
 isothermal sphere model), the deflection potential is given by
 $\psi(\mathbf{x})=\psi_0 x$ ($\psi_0$ is a constant) from
 Eq.(\ref{apsi}).
Inserting this potential $\psi$ into Eq.(\ref{ampf}), we obtain,
\beq
  F(w,y)=\frac{e^{i w [y^2 / 2 + \phi_m(y)]}}{2 \pi i w} \int d
 x^{\prime 2}
 \exp \left[ -i \left\{ y x^{\prime}_1 + \psi_0 \sqrt{x_1^{\prime 2}
 + x_2^{\prime 2}} +\mathcal{O} (1/w) \right\} \right],  
\label{dfcusp}  
\eeq
where $\mathbf{x}^{\prime}=w \mathbf{x}$. 
Denoting $dF_c(w,y)$ as the leading term of the above integral
 which is proportional to $1/w$, 
 we obtain $dF_c$ as, 
\beqa
  dF_c(w,y) &=& \frac{e^{i w [y^2 / 2 +\phi_m(y)]}}{w} \frac{1}{(y^2 -
 \psi_0^2)^{3/2}} ~~~~{\mbox{for}}~~ |y| > |\psi_0|, 
\label{dfcsis0}   \\
       &=& \frac{e^{i w [y^2 / 2+\phi_m(y)]}}{w} \frac{i}{(\psi_0^2 -
 y^2)^{3/2}} ~~~~{\mbox{for}}~~ |y| < |\psi_0|.
\label{dfcsis}
\eeqa 
Thus, the contributions to the amplification factor $F$ at the lens
 center is of the order of $1/w \sim \lambda/M$.
This is because of the singularity in the density profile at
 the lens center.
The correction terms $dF_c$ in Eqs.(\ref{dfcsis0}) and
 (\ref{dfcsis}) represent a diffracted image which is
 formed at the lens center by the diffraction effect. 
The magnification of this image is of the order of
 $\sim \lambda/M$.

\paragraph{Case of $\mbox{\boldmath{$\alpha =1$}}$}  

If the inner density profile is $\rho \propto r^{-1}$ (e.g. the Navarro
 Frenk White model), the deflection potential is given by
 $\psi(\mathbf{x})=\psi_0 x^2 \ln x$ ($\psi_0$ is a constant) from
 Eq.(\ref{apsi}).
Inserting this $\psi$ into Eq.(\ref{axiampf}), we have,
\beq
 F(w,y)=- \frac{i}{w} ~e^{i w [y^2 / 2+\phi_m(y)]} \int_0^{\infty} d
 x^{\prime} x^{\prime} J_0(y x^{\prime}) ~e^{i/(2w) x^{\prime 2} [1-2 \psi_0
 \ln (x^{\prime}/w)]}, 
\label{dfcusp2}
\eeq 
where $x^{\prime}=w x$. 
By expanding the integrand of Eq.(\ref{dfcusp2}) in powers of $1/w$,
 the above equation can be integrated analytically.
Denoting $dF_c(w,y)$ as the leading term, we obtain,
\beq
  dF_c(w,y) = \frac{-4 \psi_0}{(w y^2)^2} ~e^{i w [y^2 / 2+\phi_m(y)]}.
\eeq 
Thus, the diffracted image is formed at the lens center with 
 the magnification $\sim (\lambda/M)^2$.

\paragraph{The other cases $\mbox{\boldmath{$(0< \alpha <1$,}}$
  $\mbox{\boldmath{$1< \alpha <2)$}}$}

The deflection potential is $\psi(\mathbf{x})=\psi_0 ~x^{-\alpha+3}$
 ($\psi_0$ is a constant) from Eq.(\ref{apsi}).
Inserting $\psi$ into Eq.(\ref{axiampf}), 
 we obtain $dF_c(w,y)$ as,
\beq
  dF_c(w,y) = -\frac{1}{2} \left( \frac{y}{2} \right)^{\alpha-5} \psi_0
  w^{\alpha-3} ~e^{i w [y^2 / 2+\phi_m(y)]}~ \frac{\Gamma((5-\alpha)/2)}
 {\Gamma((\alpha-3)/2)},
\label{dfcusp3-2}
\eeq 
where $\Gamma$ is the gamma function.  \\

From the above discussion for the three cases, the diffracted image is always
 formed at the lens center for the inner density profile
 $\rho \propto r^{-\alpha}$ ($0 < \alpha \leq 2$).
The magnification of this central image is roughly given by,
 $\mu \sim (\lambda/M)^{3-\alpha}$.

\section{Results for Specific Lens Models}

We apply the quasi-geometrical optics approximation to the cases
 of the simple lens models.
We consider the following axially symmetric lens models: 
 point mass lens, singular isothermal sphere (SIS) lens,
 isothermal sphere lens with a finite core,
 and Navarro Frenk White (NFW) lens. 
We derive the amplification factor $F$, that in the geometrical optics
 limit $F_{geo}$, and its correction terms
 $dF_m$ (in \S 4.1) and $dF_c$ (in \S 4.3) for the above lens
 models.
  We define $dF$ for convenience as the sum of $dF_m$ and $dF_c$:
 $dF(w,y) \equiv dF_m(w,y) + dF_c(w,y)$.

\subsection{Point Mass Lens}

The surface mass density is expressed as
 $\Sigma(\bfxi) = M \delta^2 (\bfxi)$, where $M$ is the lens mass, 
 and the deflection potential is $\psi(\mathbf{x})= \ln x$.
This lens model is most frequently used in wave optics in
 gravitational lensing of gravitational waves (Nakamura \cite{n98};
 Ruffa \cite{r99}; De Paolis \etal \cite{dp01}, \cite{dp02};
 Zakharov \& Baryshev \cite{zb02}; Takahashi \& Nakamura \cite{tn03};
 Yamamoto \cite{y03}; Varvella \etal \cite{v03}; Seto \cite{s04}).
The amplification factor $F$ of Eq.(\ref{axiampf}) is
 analytically integrated as (Peters \cite{p74};
 Deguchi \& Watson \cite{dw86a}),
\beq
  F(w,y) = \exp\left[ \frac{\pi w}{4} + i \frac{w}{2}
 \left\{ \ln \left(  \frac{w}{2} \right) - 2 \phi_m(y) \right\} \right]
 ~\Gamma \left( 1- \frac{i}{2} w \right) ~_1 F_1 \left(
 \frac{i}{2} w,1;\frac{i}{2} wy^2 \right),
\label{afp}
\eeq
where $_1 F_1$ is the confluent hypergeometric function and
 $\phi_m(y)=(x_m-y)^2/2-\ln{x_m}$ with $x_m=(y+\sqrt{y^2+4})/2$.
In the geometrical optics limit $(w \gg 1)$ from Eq.(\ref{fgeo})
 we have,
\beq
  F_{geo}(w,y) = \left| \mu_{+} \right|^{1/2}
 -i \left| \mu_{-} \right|^{1/2} e^{i w \Delta T},
\label{gpf}  
\eeq
where the magnification of each image is $\mu_{\pm}=1/2 \pm (y^2+2)
 /(2 y \sqrt{y^2+4})$ and the time delay between double images is 
 $\Delta T = y \sqrt{y^2+4}/2 + \ln
 ((\sqrt{y^2+4}+y)/(\sqrt{y^2+4}-y))$.
In the quasi-geometrical optics approximation,
 $dF_c=0$ and $dF(=dF_m)$ is given from Eq.(\ref{dfm}) by,
\beq
  dF(w,y) = \frac{i}{3 w} \frac{4 x_+^2 -1}{(x_+^2+1)^3 (x_+^2-1)}
 \left| \mu_{+} \right|^{1/2}
 + \frac{1}{3 w} \frac{4 x_-^2 -1}{(x_-^2+1)^3 (x_-^2-1)}
 \left| \mu_{-} \right|^{1/2} e^{i w \Delta T}
\label{dgpf}
\eeq
where $x_{\pm}=(y \pm \sqrt{y^2+4})/2$ is the position of each image.
The first and second terms in Eq.(\ref{dgpf}) are correction terms for
 the magnifications of the two images as discussed in \S 4.1.

\begin{figure}
\centering
\resizebox{8.cm}{!}{\includegraphics{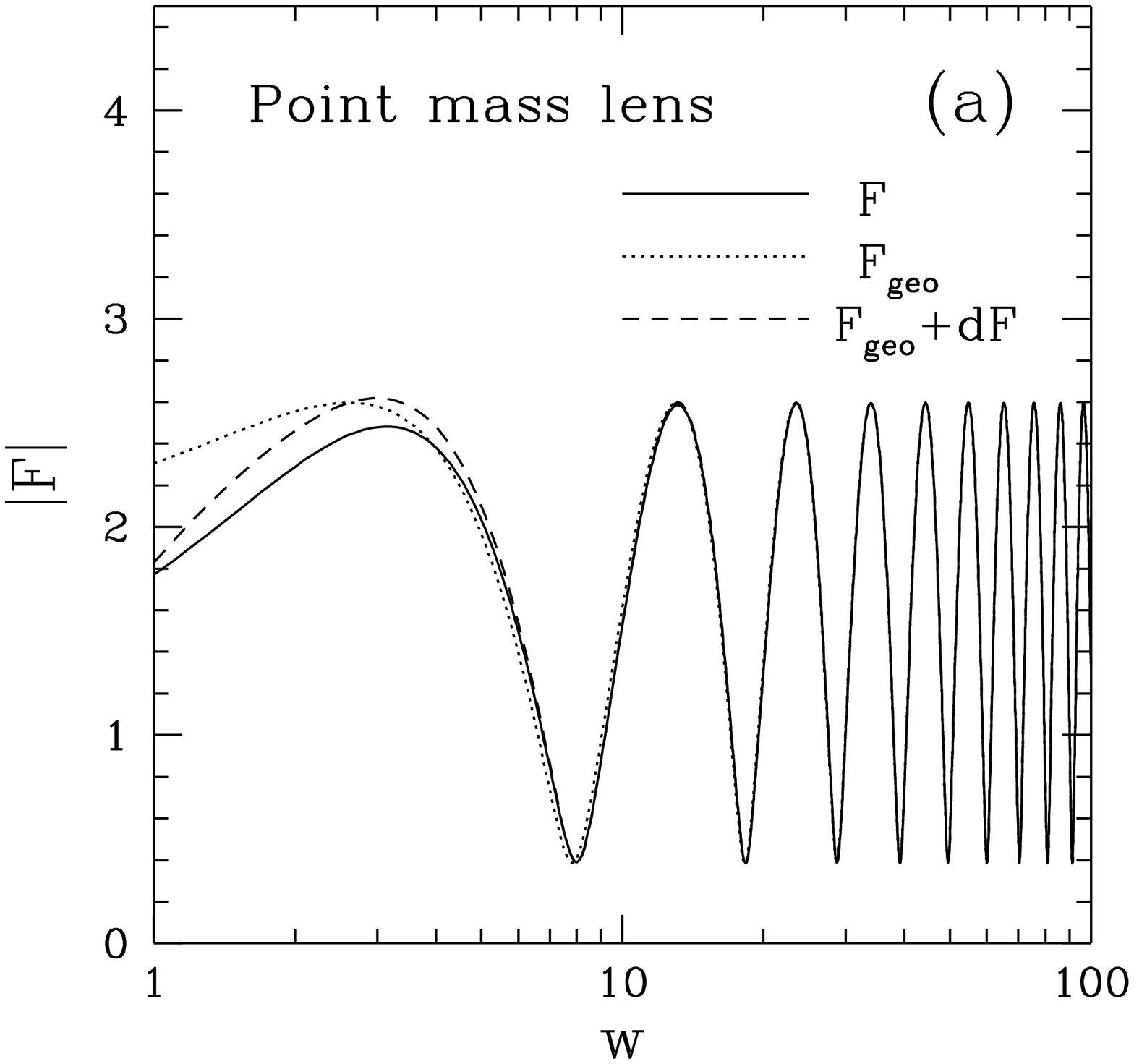}}
\hspace{1.cm}
\resizebox{8.cm}{!}{\includegraphics{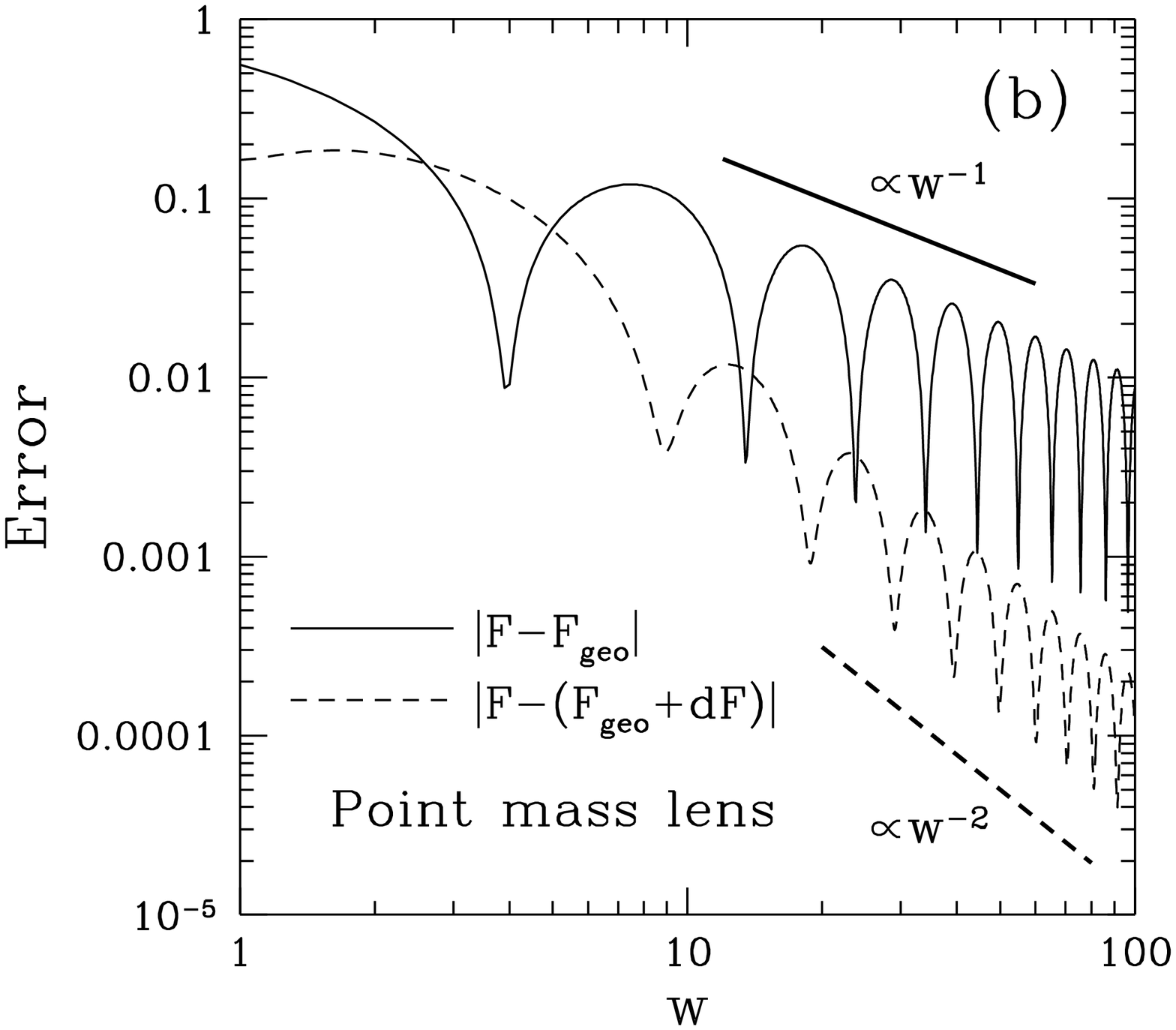}}
\caption{
(a) The amplification factor $|F(w,y)|$ for a point mass lens
 as a function of $w$ with a fixed source position $y=0.3$.
The solid line is the full result $F$; the dotted line
 is the geometrical optics approximation $dF_{geo}$;
 the dashed line is the quasi-geometrical optics approximation 
 $F_{geo}+dF$.
(b) The differences between $F$, $F_{geo}$ and 
 $F_{geo}+dF$ for a point mass lens as a function of $w$ with $y=0.3$.
The thin solid line is $|F-F_{geo}|$, and the thin dashed line
 is $|F-(F_{geo}+dF)|$.
The thick solid (dashed) line is the power of $1/w ~(1/w^2)$.}
\label{p23}
\end{figure}

In Fig.\ref{p23}(a), the amplification factor $|F|$ is shown
 as a function of $w$ with a fixed source position $y=0.3$.
The solid line is the result $F$ in Eq.(\ref{afp}); the dotted line
 is the geometrical optics approximation $F_{geo}$ in Eq.(\ref{gpf});
 the dashed line is the quasi-geometrical optics approximation 
 $F_{geo}+dF$ in Eqs.(\ref{gpf}) and (\ref{dgpf}).
For the high frequency limit $w \gg 1$, $|F|$ converge to the
 geometrical optics limit in Eq.(\ref{gpf}),
\beq
  \left| F_{geo} \right|^2 = \left| \mu_{+} \right| + \left| \mu_{-} \right| 
  + 2 \left| \mu_{+} \mu_{-} \right|^{1/2} \sin ( w \Delta T).
\label{absgpf}
\eeq
The first and second terms in above equation (\ref{absgpf}), $ | \mu | =
 | \mu_{+} | + | \mu_{-} |$, represent the total magnification
 in the geometrical optics. 
The third term expresses the interference between the double images.
The oscillatory behavior (in Fig.\ref{p23}(a)) is due to this
 interference.
The amplitude and the period of this oscillation are approximately
 equal to $2 |\mu_+ \mu_-|^{1/2}$ and $w \Delta T$ in the third 
 term of Eq.(\ref{absgpf}), respectively.
For large $w$ ($\gsim 10$), these three lines asymptotically converge.

In Fig.\ref{p23}(b), the differences between $F$, $F_{geo}$ and 
 $F_{geo}+dF$ are shown as a function of $w$ with $y=0.3$.
The thin solid line is $|F-F_{geo}|$, and the thin dashed line
 is $|F-(F_{geo}+dF)|$.
The thick solid (dashed) line is the power of $w^{-1} (w^{-2})$.
From this figure, for larger $w$ ($\gg 1$) $F$
 converges to $F_{geo}$ with the error of $\mathcal{O} (1/w)$ and
 converges to $F_{geo}+dF$ with the error of $\mathcal{O} (1/w^2)$.
These results are consistent with the analytical calculations in \S
 3 and 4.

\subsection{Singular Isothermal Sphere}

The surface density of the SIS (Singular Isothermal Sphere) is characterized
 by the velocity dispersion $v$ as, $\Sigma(\bfxi)=v^2/(2 \xi)$. 
 and the deflection potential is $\psi(\mathbf{x})=x$.
The SIS model is used for more realistic lens objects than the point
 mass lens, such as galaxies, star clusters and dark halos
 (Takahashi \& Nakamura \cite{tn03}).
In this model $F$ is numerically computed in Eq.(\ref{axiampf}).
In the geometrical optics limit $w \gg 1$, 
 $F_{geo}$ is given by,
\beqa
  F_{geo}(w,y) &=& \left| \mu_{+} \right|^{1/2}
 -i \left| \mu_{-} \right|^{1/2} e^{i w \Delta T} 
 ~~~~~~~\mbox{for}~y < 1,  \nonumber \\
       &=& \left| \mu_{+} \right|^{1/2}
 ~\hspace{2.67cm}  \mbox{for}~y \geq 1,    
\label{gsf}
\eeqa
where $\mu_{\pm}=\pm 1 +1/y$ and $\Delta T=2 y$. 
For $y < 1$ double images are formed, while for $y \geq 1$ single
 image is formed.
In quasi-geometrical optics approximation, $dF$ is given by,
\beqa
 dF(w,y) &=& \frac{i}{8 w} \frac{|\mu_{+}|^{1/2}}{y (y+1)^2}
 - \frac{1}{8 w} \frac{|\mu_{-}|^{1/2}}{y (1-y)^2} ~e^{i w \Delta T}
 + \frac{i}{w} \frac{1}{(1 - y^2)^{3/2}} ~e^{i w [y^2 / 2 + \phi_m(y) ]}
 ~\hspace{0.5cm} \mbox{for}~y < 1,   \label{dgsf}  \\
       &=& \frac{i}{8 w} \frac{|\mu_{+}|^{1/2}}{y (y+1)^2}
 + \frac{1}{w}\frac{1}{(y^2 - 1)^{3/2}} ~e^{i w [y^2 / 2 + \phi_m(y) ]} 
 ~\hspace{3.52cm}  \mbox{for}~y \geq 1,
\label{dgsf3}    
\eeqa
where $\phi_m(y)=y+1/2$.
For $y < 1$, the first and the second term in Eq.(\ref{dgsf}) are
 correction terms for the magnifications of the images formed in the
 geometrical optics (i.e. $dF_m$ in \S 4.1), and the third term corresponds  
 to the diffracted image at the lens center (i.e. $dF_c$ in \S 4.3).
For $y \geq 1$, the first term of Eq.(\ref{dgsf3}) is correction term
 for the magnification $dF_m$, and the second term corresponds  
 to the diffracted image at the lens center $dF_c$.
Thus, in the quasi-geometrical optics approximation, for $y < 1$
 the three images are formed, while for $y \geq 1$ the double images
 are formed in the SIS model. 

\begin{figure}
\centering
\resizebox{8.cm}{!}{\includegraphics{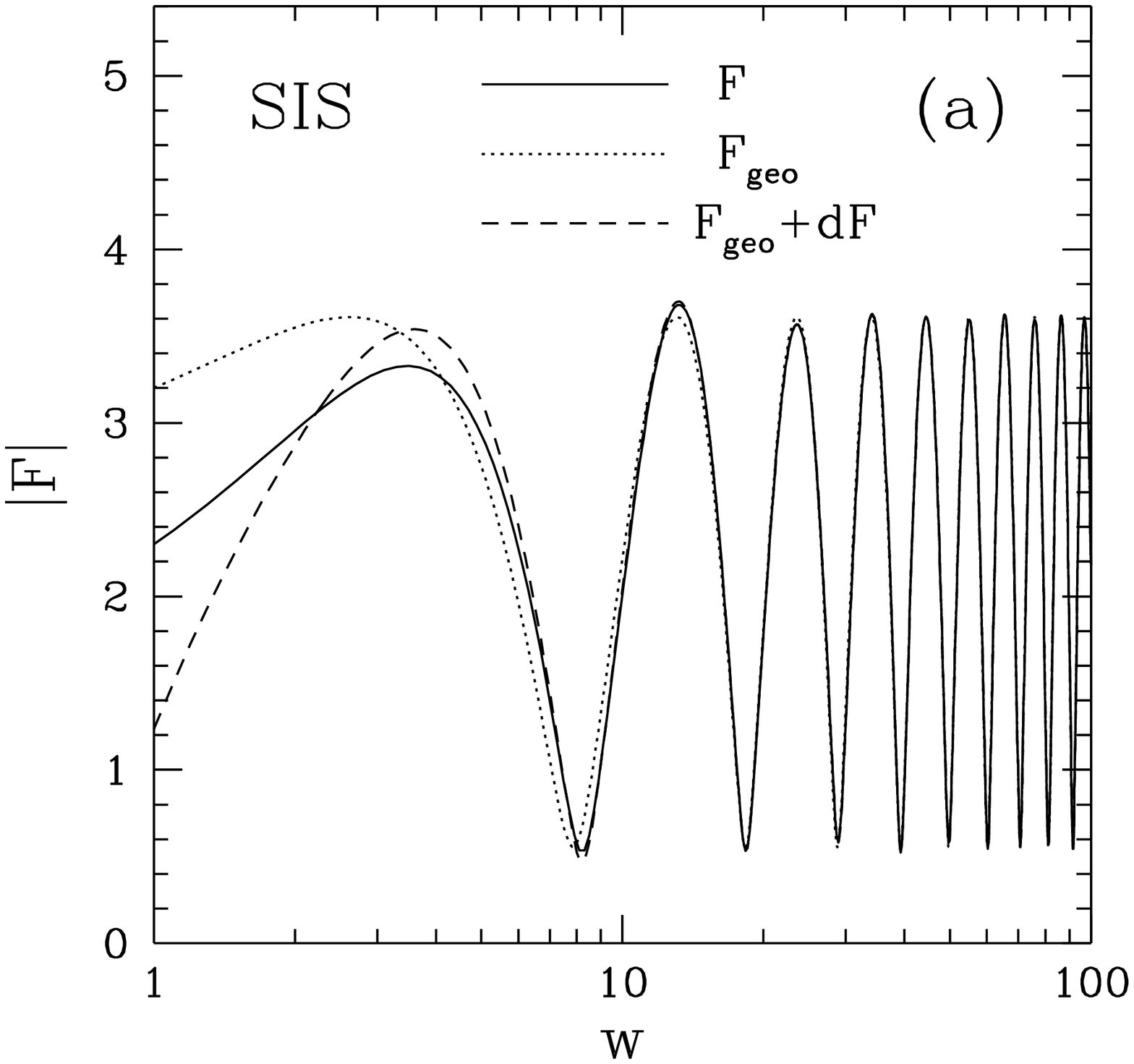}}
\hspace{1.cm}
\resizebox{8.cm}{!}{\includegraphics{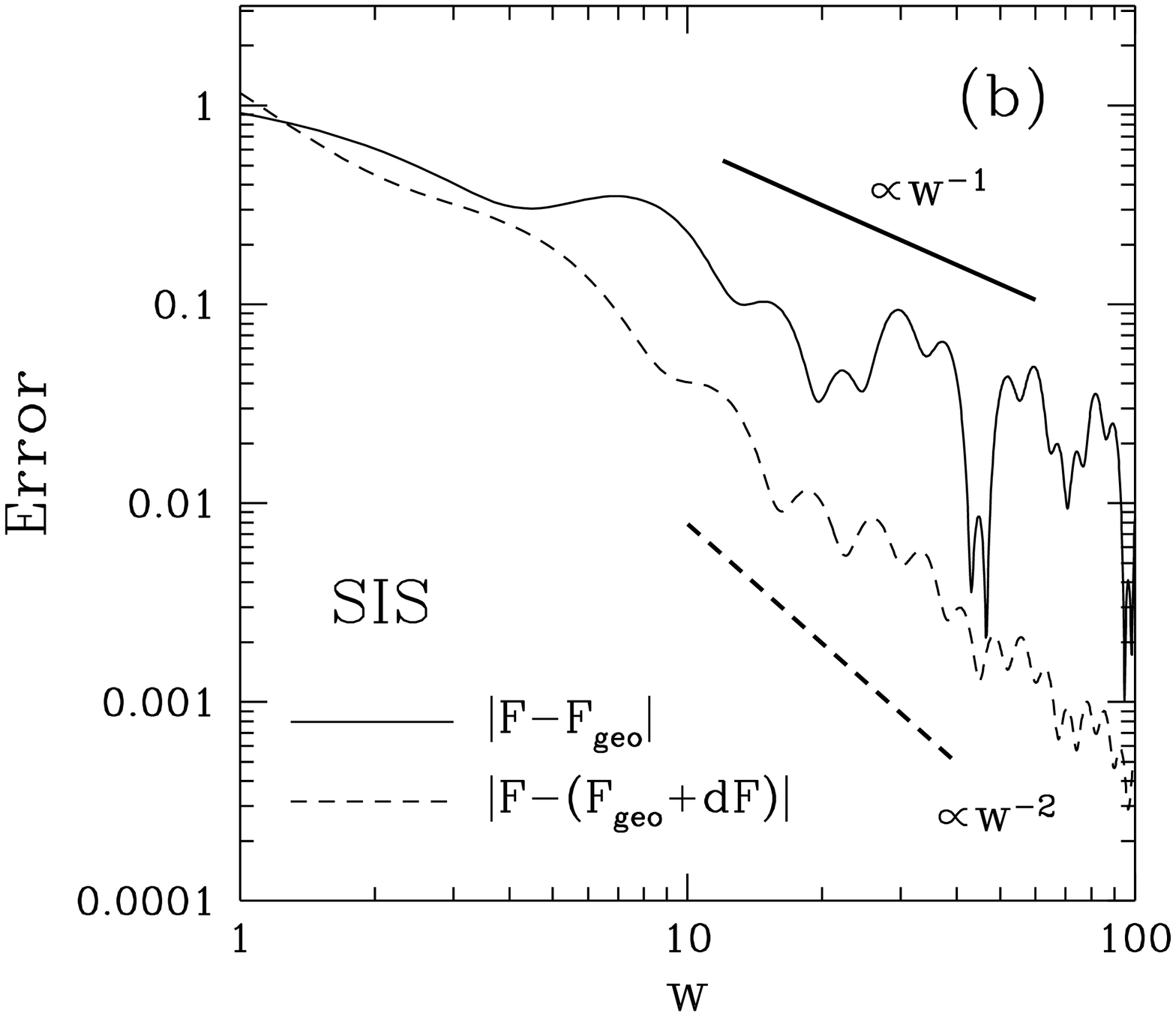}}
\caption{
Same as Fig.\ref{p23}, but for SIS lens with a source position $y=0.3$.}
\label{f23}
\end{figure}

Fig.\ref{f23} is the same as Fig.\ref{p23}, but for the SIS lens
 with a source position $y=0.3$.
In Fig.\ref{f23}(a), the behavior is similar to that in the point mass
 lens (in Fig.\ref{p23}(a)).
The oscillation of $|F_{geo}|$ is because of the interference between the
 double images, while the oscillation of $|F_{geo}+dF|$ is because of that
 among the three images.
As shown in Fig.\ref{f23}(b), the errors decrease as $w$ increases, and 
 the results are consistent with the analytical calculations.

\subsection{Isothermal Sphere with a Finite Core}

We investigate the effect of a finite core at lens center on the
 amplification factor. 
We consider the isothermal sphere having a finite core.
The deflection potential is $\psi(x)=(x^2+x^2_c)^{1/2}$ where $x_c$ is a
 dimensionless core radius.
In this model, the central core of the lens contributes the integral of
 $F$ in Eq.(\ref{ampf}).
Denoting $dF_c$ as 
 the contribution of $F$ at the lens center, we obtain,
\beq
  dF_c(w,y)=\frac{e^{i w y^2 / 2}}{2 \pi i w} \int d \mathbf{x^{\prime}}^2
 \exp \left[ -i \left\{ y x^{\prime}_1 + \sqrt{x_1^{\prime 2}
 + x_2^{\prime 2}+ (w x_c)^2} +\mathcal{O} (1/w) \right\} \right].
\label{dfcisc}
\eeq
For $w x_c \lsim 1$ the above equation is the same as $dF_c$ in the SIS
 model (see Eq.(\ref{dfcusp})) and $dF_c \propto 1/w$, but for
 $w x_c \gsim 1$ $dF_c$ exponentially decreases as $w$ increases.
Thus even if the lens has the finite core at the center, the wave 
 does not feel the existence of the small core $x_c \lsim 1/w$, and
 the diffracted image is formed similar to the lens without the core.

\subsection{NFW lens}

The NFW profile was proposed from numerical simulations of cold dark
 matter (CDM) halos by Navarro, Frenk \& White (\cite{nfw97}).
They showed that the density profile of the dark halos has the
 ``universal'' form,
 $\rho (r) = \rho_s (r/r_s)^{-1} (r/r_s+1)^{-2}$,
where $r_s$ is a scale length and $\rho_s$ is a characteristic density.
The NFW lens is used for lensing by galactic halos and clusters of
 galaxies.
The deflection potential is written as (Bartelmann \cite{b96}),
\beqa
  \psi(x) &=& \frac{\kappa_s}{2} \left[ \left( \ln \frac{x}{2} \right)^2
  - \left(  \mbox{arctanh} \sqrt{1-x^2} \right)^2  \right] 
 ~~~\mbox{for} ~~x \le 1, \nonumber \\
  &=& \frac{\kappa_s}{2} \left[ \left( \ln \frac{x}{2} \right)^2
  + \left(  \arctan \sqrt{x^2-1} \right)^2  \right] 
 ~\hspace{0.3cm} \mbox{for} ~~x \ge 1, 
\label{psinfw}
\eeqa
where $\kappa_s = 16 \pi \rho_s (D_L D_{LS}/D_S) r_s$ is the 
 dimensionless surface density and we adopt the scale radius,
 not the Einstein radius, as the normalization length: $\xi_0=r_s$. 
The image positions $x_j$, magnifications $\mu_j$ and time delays $T_j$
 are numerically obtained from the lens equation, $y=x-\psi^{\prime}(x)$. 
In the quasi-geometrical optics approximation, $dF$ is given by,
\beq
 dF(w,y)=\frac{i}{w} \sum_j  \Delta_j 
 \left| \mu_j \right|^{1/2}  e^{i w T_{j}- i \pi n_j} + \frac{\kappa_s}
 {(wy^2)^2}~e^{iw[y^2/2+\phi_m(y)]}.
\label{dfnfw}
\eeq
The first term in Eq.(\ref{dfnfw}) is the corrections for the
 magnifications of the images (i.e. $dF_m$ in \S 4.1).
The second term corresponds to the diffracted image at the lens center
 (i.e. $dF_c$ in \S 4.3).

\section{Conclusions}

We studied the gravitational lensing in the quasi-geometrical optics
 approximation which is the geometrical optics including the corrections
 arising from the effect of the finite wavelength.
Theses correction terms can be obtained analytically by the asymptotic 
 expansion of the diffraction integral in powers of the wavelength
 $\lambda$.
The first term, arising from the short wavelength limit
 $\lambda \to 0$, corresponds to the geometrical optics limit.
The second term, being of the order of $\lambda/M$ ($M$ is the
 Schwarzschild radius of the lens), is the first correction term
 arising from the diffraction effect.
By analyzing this correction term, we obtain the following results:
 (1)The lensing magnification $\mu$ is modified to
       $\mu ~(1+\delta)$, where $\delta$ is of the order of
       $(\lambda/M)^2$.
 (2)If the lens has cuspy (or singular) density profile at the center
 $\rho(r) \propto r^{-\alpha}$ ($0 < \alpha \leq 2$) the diffracted image
 is formed at the lens center with the magnification
 $\mu \sim (\lambda/M)^{3-\alpha}$. 
 Thus if we observe this diffracted image by the various wavelength 
 (e.g. the chirp signal), the slope $\alpha$ can be determined.

\begin{acknowledgements}
 I would like to thank Takashi Nakamura, Takeshi Chiba, and
 Kazuhiro Yamamoto for useful comments and discussions.

\end{acknowledgements}

\appendix

\section{Numerical Computation for the Amplification Factor}

We present the method for the numerical integration of $F$ discussed
 in \S 2. $F$ in Eq.(\ref{ampf}) is rewritten as,
\beq
  F(w,\mathbf{y}) = \frac{w}{2 \pi i} ~e^{iw[y^2/2+\phi_m(\mathbf{y})]}
 \int_0^{\infty} dx~x~e^{iwx^2/2} \int_0^{2 \pi} d \theta
 ~e^{-iw [xy \cos \theta + \psi (x,\theta)]},
\label{appc7}
\eeq
where $\theta$ is defined as $\mathbf{x} \cdot \mathbf{y} = xy \cos
 \theta$.
Changing the integral variable from $x$ to $z=x^2/2$ in
 Eq. (\ref{appc7}), we obtain,
\beq
 F(w,\mathbf{y}) = \int_0^{\infty} dz f(z:w,\mathbf{y}) ~e^{i w z}
\label{appc4}
\eeq
where the function $f$ is defined by,
\beq
 f(z:w,\mathbf{y}) \equiv \frac{w}{2 \pi i}
 ~e^{iw[y^2/2+\phi_m(\mathbf{y})]} \int_0^{2 \pi} d \theta
 ~e^{-iw [y \sqrt{2 z} \cos \theta + \psi (\sqrt{2 z},\theta)]}.
\label{appc9}
\eeq
Thus, we obtain the Fourier integral (\ref{appc4}).
We present the method for computing Fourier integral in Numerical
 Recipes (Press \etal \cite{p92}).
The equation (\ref{appc4}) is rewritten as,
\beqa
  F(w,y) &=& \int_0^{b} dz f(z:w,y) ~e^{i w z} + 
             \int_b^{\infty} dz f(z:w,y) ~e^{i w z},  \nonumber \\
 &=& \int_0^{b} dz f(z:w,y) ~e^{i w z} - \frac{f(b:w,y) e^{iwb}}{iw}
        + \frac{f^{\prime}(b:w,y) e^{iwb}}{(iw)^2} - \cdot \cdot \cdot,
\label{appc6}
\eeqa
where $f^{\prime} = \partial f / \partial_{}b$.
The first term of the above equation (\ref{appc6}) is evaluated by
 numerical integration directly.
For the series of the above equation, we use the integration by parts.
The asymptotic expansion in Eq.(\ref{appc6}) converges for large $b$.

Especially for the axially symmetric lens, the function $f$ in
 Eq.(\ref{appc9}) is reduced to the simple form as,
\beq
  f(z:w,y)=-iw e^{iw [y^2/2+\phi_m(y)]} J_0 (wy \sqrt{2z})
   ~e^{iwz}.
\eeq

\section{Asymptotic Expansion of the Amplification Factor for
 Non-axially Symmetric Lens Models}

We consider the expansion of $F$ in powers of $1/w$
 for the non-axially symmetric lens model.
In this case, the expansion of $T(\mathbf{x},\mathbf{y})$ around the
 j-th image position $\mathbf{x}_j$ in Eq.(\ref{expt2}) is rewritten as,
\beq
  T(\mathbf{x},\mathbf{y}) = T_j  + \frac{1}{2} \sum_{a,b} 
 T_{ab} \tilde{x}_a \tilde{x}_b + \frac{1}{6} \sum_{a,b,c} \partial_a
 \partial_b \partial_c T(\mathbf{x}_j,\mathbf{y}) \tilde{x}_a
 \tilde{x}_b \tilde{x}_c  + \frac{1}{24} \sum_{a,b,c,d} \partial_a
 \partial_b \partial_c \partial_d T(\mathbf{x}_j,\mathbf{y})
 \tilde{x}_a \tilde{x}_b \tilde{x}_c \tilde{x}_d
 + \mathcal{O} (\tilde{x}^5),
\label{appi1}
\eeq
where $T_{ab} = \partial_a \partial_b T(\mathbf{x}_j,\mathbf{y})$
 is a $2 \times 2$ matrix.  
We change the variable from $\tilde{x}_a$ to $z_a = \sum_b A_{ab}
 \tilde{x}_b$ in order to diagonalize the matrix $T_{ab}$ in the 
 above equation (\ref{appi1}).
Here, $A_{ab}$ satisfies,
  $\sum_{a,b} T_{ab} A_{ac} A_{bd} = \lambda_c \delta_{cd}$, 
where $\lambda_c$ is the eigenvalue of $T_{ab}$. 
Using the variable $z$, the equation (\ref{appi1}) is rewritten as,
\beqa
   T(\mathbf{x},\mathbf{y}) = T_j + \frac{1}{2} \left( \lambda_1 z_1^2
 + \lambda_2 z_2^2 \right) + \sum_{a,b,c} M_{abc} ~z_a z_b z_c 
 + \sum_{a,b,c,d} N_{abcd} ~z_a z_b z_c z_d + \mathcal{O} (z^5), 
\label{appi5}
\eeqa
where $M_{abc}$ and $N_{abcd}$ are defined by,
\begin{eqnarray*}
  M_{abc} &\equiv& \frac{1}{6} \sum_{d,e,f} \partial_d \partial_e
 \partial_f T(\mathbf{x}_j,\mathbf{y}) ~A_{ad} A_{be}
 A_{cf}, \nonumber \\
  N_{abcd} &\equiv& \frac{1}{24} \sum_{e,f,g,h} \partial_e \partial_f
 \partial_g \partial_h T(\mathbf{x}_j,\mathbf{y}) ~A_{ae} A_{bf}
 A_{cg} A_{dh}.
\end{eqnarray*}

Doing the same calculations in \S 4.1 with Eq.(\ref{appi5}), we obtain
 the amplification factor same as equation (\ref{fdgeo}), but
 $\Delta_j$ is given by
\beqa
  \Delta_j &=& \frac{15}{2} \left( \frac{M_{111}^2}{\lambda_1
  |\lambda_1|^2} + 3 \frac{M_{111} M_{122}}{|\lambda_1|^2 \lambda_2}
 + 3 \frac{M_{112} M_{222}}{\lambda_1 |\lambda_2|^2}
 + \frac{M_{222}^2}{\lambda_2 |\lambda_2|^2} \right)
 -3 \left( \frac{N_{1111}}{|\lambda_1|^2} 
 + 2 \frac{N_{1122}}{\lambda_1 \lambda_2}
 + \frac{N_{2222}}{|\lambda_2|^2} \right).
\label{appi9}
\eeqa

\end{document}